\documentclass[final,3p,times,twocolumn]{elsarticle}

\usepackage{amsmath}
\usepackage{amssymb}
\usepackage{graphicx}

\journal{Physica E}

\newcommand{\cE}{{\epsilon}}
\newcommand{\Vdc}{{V_{\rm dc}}}
\newcommand{\Vac}{{V_{\rm ac}}}
\newcommand{\cS}{{\cal S}}

\newcommand{\acom}[2]{{\{ #1 , #2 \}}}

\newcommand{\vG}{\check G}
\newcommand{\vgt}{\check \tau}

\newcommand{\Tr}{{\rm Tr}}

\begin{document}

\begin{frontmatter}



\title{Elementary Andreev Processes in a Driven\\
Superconductor-Normal Metal Contact}


\author[1]{Wolfgang Belzig}

\ead{Wolfgang.Belzig@uni-konstanz.de}

\address[1]{Fachbereich Physik, Universit\" at Konstanz,
D-78457 Konstanz, Germany}

\author[2]{Mihajlo Vanevic}

\ead{mihajlo.vanevic@gmx.com}

\address[2]{Department of Physics, University of Belgrade,
11158 Belgrade, Serbia}

\begin{abstract}

We investigate the full counting statistics of a voltage-driven normal
metal(N)-superconductor(S) contact. In the low-bias regime below the
superconducting gap, the NS contact can be mapped onto a purely normal
contact, albeit with doubled voltage and counting fields. Hence in
this regime the transport characteristics can be obtained by the
corresponding substitution of the normal metal results. The elementary
processes are single Andreev transfers and electron- and hole-like
Andreev transfers. Considering Lorentzian voltage pulses we find an
optimal quantization for half-integer Levitons. 

\end{abstract}

\begin{keyword}

Quantum transport\sep
Andreev reflection\sep
Time-dependent drive\sep
Full counting statistics




\end{keyword}

\end{frontmatter}


\section{Introduction}
\label{sec:intro}

Quantum shot noise and full counting statistics (FCS) have emerged as
central tools of quantum transport in the last two decades. The main
driving force is the dramatic difference in properties of fluctuations
of the current of classical particles versus quantum particles behaving
sometimes in a wave-like fashion \cite{BlanterButtikerPHYSREP2000}.
Classical particles in a tunneling setup lead to fluctuations in the
current described by Schottky's formula for Poisson noise $S=eI$,
where $S$ is the noise power of current fluctuations, $I$ is the
average current and $e$ is the electron charge \cite{SchottkyANNPHYS1918}.
Considering the wave-like nature of electrons, which is encountered
in nanostructured conductors at low temperatures, in combination with
the quantum statistical fermionic Pauli principle leads to a
suppression of the shot noise by a so called Fano factor
$F=\sum_n T_n(1-T_n)/\sum T_n$, where $T_n$ are the transmission
probabilities of the electron waves in channels
$n$ \cite{KhlusJETP87,LesovikJETPLETT89}.
The suppression has been experimentally verified
in quantum point contacts \cite{ReznikovPRL75-95,GlattliPRL76-96}
and other coherent conductors like
diffusive wires with the characteristic Fano factor $F=1/3$
\cite{BeenakkerButtikerPRB46-92,DevoretPRL76-96,SchonenbergerPRB59-99}.

Another leap forward in the understanding of quantum transport was
to go beyond the average current and the noise by considering the full
counting statistics (FCS) of the transferred charge, which comprises
all probabilities $P(N)$ to transfer $N$ charges. Equivalently, one
considers the cumulant generating function (CGF)
$\cS(\chi)=\ln[ \langle e^{i\chi N} \rangle ]$. The remarkable result for
a quantum contact at low temperature is
$\cS(\chi)=(2eVt_0/h)\sum_n\ln\left[1+T_n(e^{i\chi}-1)\right]$
which describes a binomial distribution for each channel \cite{LevitovLesovikJETPLETT58-93}.
Note that in general a decomposition of the CGF into binomials or
multinomials allows to identify the elementary processes and their
probabilities.

In superconductors the electrons are correlated in a single
macroscopic wave function, which describes a condensate of so-called
Cooper pairs consisting of two bound electrons. The state is
stabilized by a finite binding energy $\Delta$, which needs to be
payed twice to break up a Cooper pair into two independent
electrons. In quantum transport this is manifest in an energy gap
$\Delta$ below which the differential conductance of a junction
between a normal metal and a superconductor due to single electrons
vanishes. However, the electron transport is still possible by an
intriguing process called Andreev reflection in which a Cooper pair
is transferred into a superconductor while the hole-like quasiparticle is left
behind \cite{AndreevJETP19-64}. Hence, in this process two charges are
transferred which is therefore possible also at subgap
energies. This process occurs with the probability of Andreev
reflection $R_n^A=T_n^2/(2-T_n)^2$
\cite{LambertJPHYSCONDMAT3-91,BeenakkerPRB46-92}. It is very
interesting to note that the FCS for Andreev reflection takes a very similar
form as in the normal case, namely
$\cS_A(\chi)=(2eVt_0/h)\sum_n\ln\left[1+R^A_n(e^{i2\chi}-1)\right]$
\cite{MuzykantskiiPRB50-94}.
Therefore, the statistics is also binomial, but with the important
difference that in each process $2$ charges are transferred. This
follows from the $\pi$-periodicity due to a doubling of the counting
field $\chi$ in the factor $e^{i2\chi}-1$. The doubling of the effective charge
transported is manifest in the ratio between noise and average
current $S/I=eF_A$, where the Fano factor is now
$F_A=2\sum_n R_n^A(1-R_n^A)/\sum_n R_n^A$. In particular,
$F_A=2/3$ in a diffusive normal-metal -- superconductor junction
\cite{deJongBeenakkerPRB49-94,JehlNATURE405-00,NagaevButtikerPRB63-01}.

Time-dependent voltage drives can be used to probe the dynamics of
electrons in transport. One interesting aspect is that with a signal
with a finite frequency $\omega$ one has a tool to access the
internal time-scale of the manybody state given by $eV/\hbar$ at low
enough temperatures. This shows up, for example, in the noise of a
quantum contact driven by harmonic voltage. The noise is a piecewise
linear function of the dc voltage bias with slopes which depend on
the amplitude of the ac voltage component. The kinks in noise occur
at dc bias voltages $eV=n\hbar\omega$ matching an integer multiple of
the drive frequency \cite{LesovikLevitovPRL72-94}.
In the normal case, the noise in the presence of the drive is always larger
than or equal to the dc noise level.
The scattering theory of the excess photon-assisted noise has been
put forward by Pedersen and B{\"u}ttiker \cite{PedersenButtikerPRB58-92}.
The further advancement was interpretation of the noise and current
cross-correlations in terms of excited electron-hole pairs that was
given by Rychkov, Polianski, and B{\"u}ttiker
\cite{RychkovPolianskiButtikerPRB72-05} for an ac drive of low amplitude,
$e\Vac\ll \hbar\omega$, where at most one electron-hole pair can be created
per voltage cycle. Remarkably, this picture of electron-hole
pairs created by the drive persists even at large amplitudes
and to all orders in charge transfer statistics
\cite{VanevicNazarovBelzigPRL99-07,VanevicNazarovBelzigPRB78-08}.
Photon-assisted noise has been observed experimentally in normal coherent
conductors \cite{ReydelletGlattliPRL90-03,SchoelkopfProberPRL80-98} and in diffusive
normal metal - superconductor junctions \cite{Kozhevnikov:2000dg}.
More recently, quantum noise oscillations have been observed in a driven
tunnel junction \cite{ReuletPRB88-13}. The noise spectral density for
dc- and ac-bias voltages for normal metal-superconducting contacts has
been discussed in \cite{TorresMartinLesovikPRB63-01}.

An extremely intriguing possibility is
the ability to control the electron dynamics by shaping the voltage
pulses. In particular, it was shown that Lorentzian voltage pulses
with a quantization condition $e\int dt V(t)=nh$ result in the
soliton-like electronic excitations which minimize the noise level to
the one of an equivalent dc voltage, $S=eIF$
\cite{LeeLevitovARXIV95,IvanovLeeLevitovPRB56-97}. These
so-called levitons are hence a collective single-electron excitations localized in
space and time, which offer interesting perspectives as carriers of
quantum information \cite{Dubois:2013dv,Jullien:2014ii}.
To access the full counting statistics in the
presence of a time-dependent drive a non-equilibrium quantum field
theoretical approach to FCS was developed by Nazarov and one of the authors
\cite{BelzigNazarovPRL87-01}.
This allowed to perform the analysis of the FCS in terms of elementary
events for an arbitrary time-dependent voltage $V(t)$
\cite{VanevicNazarovBelzigPRL99-07,VanevicNazarovBelzigPRB78-08}.
The results is that one has to distinguish two
types of events: Single electron transfers, which occur with a
frequency of the average voltage and have the standard binomial statistics
$\cS_1(\chi)=(2e\Vdc t_0/h)\sum_n\ln\left[1+T_n(e^{i\chi}-1)\right]$ and
electron-hole pairs obeying a trinomial statistics
$\cS_{eh}(\chi)=\sum_{nk}M_k\ln\left[1+2T_nR_np_k(\cos(\chi)-1)\right]$.
The probabilities $p_k$ are interpreted as probabilities of electron-hole
pair creations and depend in a characteristic way on the driving
voltage which is assumed to be periodic with frequency $\omega$.
The number of attempts for the pairs to traverse the contact is
$M\propto 2\hbar\omega t_0/h$.
This opens a route towards dynamic control of elementary excitations
using suitably tailored voltage pulses \cite{VanevicBelzigPRB86-12}.

In this article, we consider the FCS of an Andreev contact driven by a
time-dependent voltage. Using an exact mapping of an NS contact onto
an effective normal contact \cite{BelzigSamuelssonEPL64-03}, we identify the
elementary Andreev events and characterize the two types of
processes. Single Andreev-pair transfers have binomial statistics and are
determined by the average dc voltage $\Vdc$. The time-dependent drive
manifests itself in correlated electron-hole pairs which are
transferred coherently. The respective probabilities are found from the
normal ones by the mapping $p_k^A[V(t)] \leftrightarrow p_k^N[2V(t)] $.
Indeed, by considering as an example the Lorentzian voltage pulses
we find a maximal noise suppression for half-integer pulses
$e\int V(t)dt=nh/2$ with integer $n$. Furthermore, increasing the
voltage level in an ac-driven contact above the gap, we find a
transition to minima at integer quantized voltages $e
V_{dc}=n\hbar\omega$.

The article is organized as follows. In Sec.~\ref{sec:keldysh}, we
introduce the extended Keldysh Greens function theory of quantum
transport applied to a time-dependent voltage drive.
In Sec.~\ref{sec:fcs}, we obtain the mapping of an NS contact
to an NN contact and analyze the resulting FCS in terms of elementary
events. Finally, in Sec. \ref{sec:ex} we discuss some examples of a
voltage drive and consider the transition from Andreev to normal
transport for large biases.

%
%

\section{Keldysh formulation of Andreev contacts}
\label{sec:keldysh}

The Keldysh Greens function formalism is a very powerful method suitable for quantum nonequilibrium problems. 
 We formally introduce the
standard closed time-path 
and define Greens functions on the contour
$G(t,t')=-i\langle \mathcal{T}_c\psi(t)\psi^\dagger(t') \rangle$
mapped onto Keldysh space. Treating the time variables on the upper
and the lower branches of the contour as independent, one can define
a matrix Greens function
\begin{equation}
  \label{eq:1}
  \hat G(t,t')=-i
  \begin{pmatrix}
    \langle {\mathcal T}\psi(t)\psi^\dagger(t')\rangle &
    \langle \psi(t)\psi^\dagger(t')\rangle \\
    \langle \psi^\dagger(t')\psi(t)\rangle &
    \langle \tilde{\mathcal T}\psi(t)\psi^\dagger(t')\rangle
  \end{pmatrix}.
\end{equation}
In the quasiclassical approximation for a free Fermi gas at
equilibrium, the Greens function reads
\begin{equation}
  \label{eq:2}
  \hat G(\epsilon)=-i\pi N_0
  \begin{pmatrix}
    1-2f(\epsilon) & -2 f(\epsilon) \\
    -2(1-f(\epsilon)) & 2 f(\epsilon)-1
  \end{pmatrix}\,
\end{equation}
The prefactor $N_0$ containing the density of states at the Fermi
level is usually removed by proper normalization, so that the Greens
function obeys the normalization condition $\hat G^2=1$. Adding
superconductivity results in the replacement
$\psi(t) \to (\psi_e(t),\psi_h(t))^T\equiv
(\psi_\uparrow(t),-\psi^\dagger_{\downarrow}(t))^T$
and hence an extension of the Keldysh matrix space by an additional
electron-hole degree of freedom (Nambu space).

For the present purpose a representation is chosen in which the
Keldysh matrices are blocks in the Nambu space. Hence, the Keldysh
Greens function of the normal lead is given by:
\begin{equation}
\check G_N =
\begin{pmatrix}
    \hat G_N^e & 0 \\
    0 & -\hat G_N^h
\end{pmatrix},
\end{equation}
where check($\check{\,\,}$) denotes matrices in Nambu($\bar{\,\,}$)
$\otimes$ Keldysh($\hat{\,\,}$) space.
Here, we introduce the (Keldysh-rotated) Greens functions
\begin{equation}
\hat G_N^e =
\begin{pmatrix}
    1 & 2UhU^\dagger \\
    0 & -1
\end{pmatrix},
\quad \hat G_N^h =
\begin{pmatrix}
    1 & 2U^\dagger h U \\
    0 & -1
\end{pmatrix},
\end{equation}
where $h=1-2f$, $f(\epsilon)=[\exp(\beta\epsilon)+1]^{-1}$ is the
Fermi distribution function, and
$U$ is related to the time-dependent drive $V(t)$,
\begin{equation}
U(t',t'') = e^{-i\int_0^{t'} eV(t)dt/\hbar}\, \delta(t'-t'').
\end{equation}

To access the FCS the counting field is incorporated into the
Green's function as
\begin{equation}
\check G_N(\chi)
=
e^{-i\chi\vgt_K/2}\, \check G_N(0)\, e^{i\chi\vgt_K/2},
\end{equation}
where $\vgt_K = \bar\tau_3 \otimes \hat\tau_1$.
This gives
\begin{equation}
\check G_N(\chi)
=
\begin{pmatrix}
    \hat G_N^e(\chi) & 0\\
    0 & -\hat G_N^h(-\chi)
\end{pmatrix}.
\end{equation}
%
The Greens function of the superconducting lead at low temperatures
and drive energies well below the gap
($k_BT_e, |eV(t)|\ll\Delta$) is given by
\begin{equation}\label{eq:GS}
\check G_S = \bar\tau_2 \otimes \hat 1
=
\begin{pmatrix}
    0 & -i\, \hat 1 \\
    i\, \hat 1 & 0
\end{pmatrix}.
\end{equation}

Cumulant generating function is given by
\cite{BelzigNazarovPRL87-01}:
\begin{equation}\label{eq:S-a}
\cS(\chi) = \frac{1}{2} \Tr\ln
\left(
    1 + \frac{T_n}{4}
    \Big(\acom{\check G_N(\chi)}{\check G_S(0)}-2\Big)
\right),
\end{equation}
where $\Tr$ stands for the trace in Nambu, Keldysh, and time (energy)
indices, and also implies a summation over transport channels $T_n$.
[$\chi$-independent constant which ensures $\cS(\chi=0)=0$ is
omitted for brevity.] Note that in deriving Eq.~(\ref{eq:S-a}) it was
assumed that the dwell time in the scattering region is very short, so
that the energy dependence of the scattering amplitudes can be neglected.

\section{Full counting statistics analysis}
\label{sec:fcs}

\begin{figure}[tb]
  \centering
  \includegraphics[width=0.9\columnwidth]{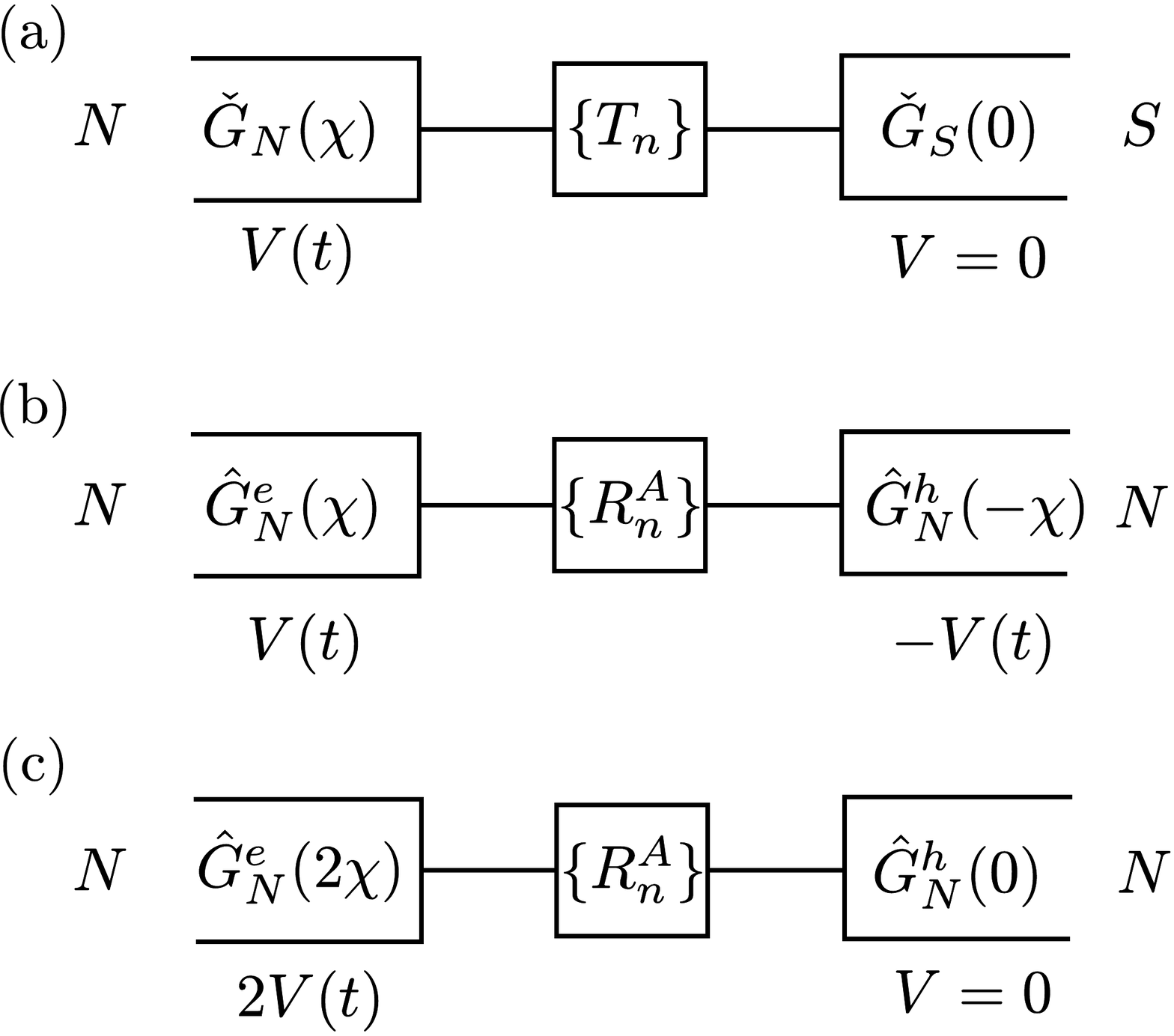}
  \caption{An Andreev contact (a) with transmission probabilities $\{T_n\}$
  between a normal and a superconducting
  metal is mapped onto a contact (b) between two normal metals
  (electron and hole space), where the transmission probabilities are
  replaced by the Andreev reflection probabilities $\{R_n^A\}$.
  (c) By a further gauge transformation the counting field and the
  (time-dependent) voltage are applied on one side only.}
  \label{fig:circuit}
\end{figure}

At low energies, when $\check G_S$ is given by Eq.~\eqref{eq:GS},
the CGF reduces to the normal-state circuit with
the electron and the hole Green's functions:
\begin{equation}
\cS(\chi) =
\frac{1}{2} \Tr\ln
\left(
    1 + \frac{R^A_n}{4}
    \Big( \acom{\hat G_N^e(\chi)}{\hat G_N^h(-\chi)} -2 \Big)
\right).
\end{equation}
Note that now $\hat G_N^{e,h}$ are the normal-state Green's
functions in the Keldysh space, and the transmission probabilities
$T_n$ are replaced by Andreev reflection probabilities,
$R^A_n = T_n^2/(2-T_n)^2$.
After carrying out a gauge transformation, it is possible to ascribe
the counting field and the drive to one 'lead' only, and we obtain
\begin{equation}\label{eq:S-c}
\cS(\chi) =
\frac{1}{2} \Tr\ln
\left(
    1 + \frac{R^A_n}{4}
    \Big( \acom{\hat G_N^e(2V(t),2\chi)}{\hat G_N^{h0}} -2 \Big)
\right),
\end{equation}
where $\hat G_N^{h0} = \hat G_N^h(V=0,\chi=0)$. Thus, at the
subgap energies, the system can be mapped to a normal-state circuit
with a doubled voltage drive and a doubled counting field,
and with transmission probabilities $T_n$ replaced by $R^A_n$.
This mapping is shown in Fig.~\ref{fig:circuit}.


CGF in Eq.~\eqref{eq:S-c} can further be brought in the form:
\begin{align}
\cS(\chi) =& \Tr\ln
\Big[
    1  + (1-\tilde f)f R^A_n (e^{2i\chi}-1) \notag
\\
& + \tilde f(1-f) R^A_n (e^{-2i\chi}-1)
\Big],
\end{align}
where
$\tilde f = U^2 \, f\, (U^\dagger)^2$
accounts for an effective doubling of the drive voltage. At zero
temperature the matrix
operators $f$, $\tilde f$ have the additional property that $f^2=f$ and
$\tilde f^2 = \tilde f$, which allows us to decompose the FCS into the
single electron processes and electron-hole pairs as mentioned in the
introduction. Here in the Andreev case they take a slightly different
form
\begin{align}
  \label{eq:S1}
  \cS_1
  &
  = \frac{2e\Vdc t_0}{h} \sum_n
  \ln\left[1+R^A_n(e^{i2\chi}-1)\right],
  \\
  \label{eq:Seh}
  \cS_{eh}
  &
  =
  M 
  \sum_{nk} \ln\left[1+2R^A_n(1-R^A_n)p_k(\cos(2\chi)-1)\right].
\end{align}

\begin{figure}[t]
\begin{center}
\includegraphics[width=0.95\columnwidth]{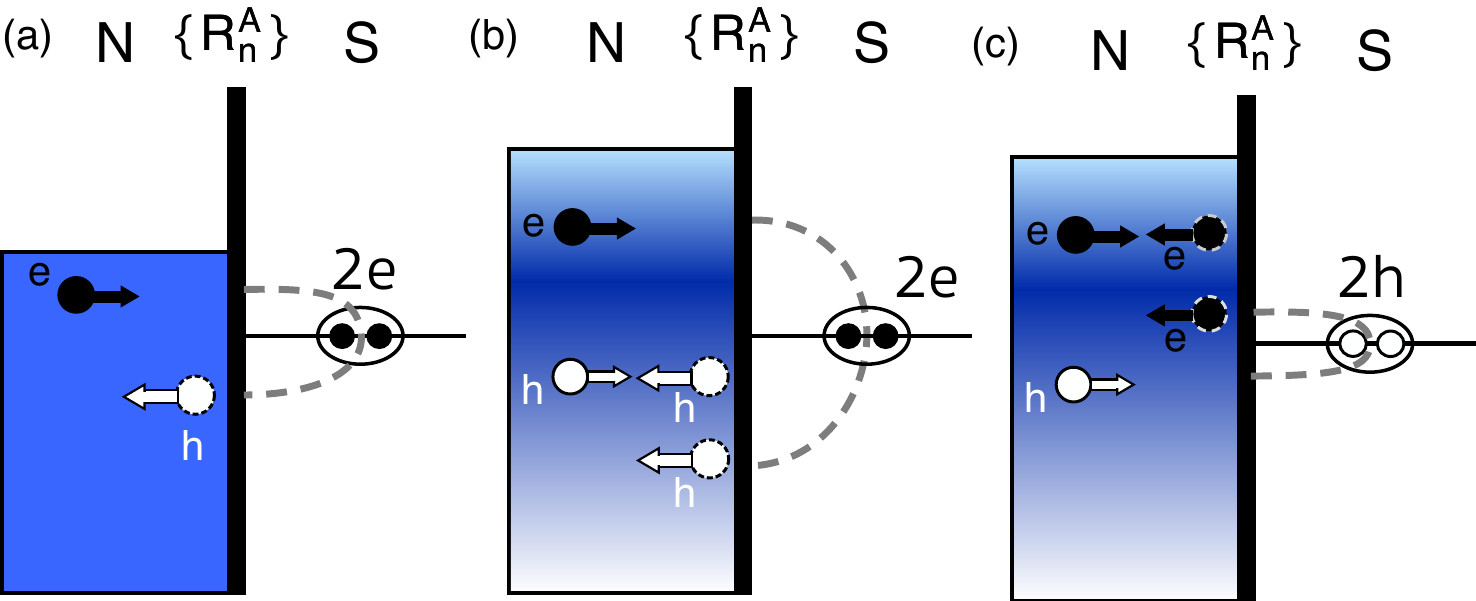}
\end{center}
\caption{Elementary transport processes in a driven NS contact:
(a) Andreev reflection of an excess electron due to dc voltage
applied, see Eq.~\eqref{eq:S1}. (b) Andreev reflection of an electron
from the electron-hole pair accompanied by the normal reflection
of the hole from the pair, see Eq.~\eqref{eq:Seh}.
A reverse process is also possible in which
the hole exhibits Andreev reflection and the electron exhibits normal
reflection. Electrons and holes are defined relative to the
energy of the superconducting condensate ($\cE=0$).
Time-dependent drive is indicated by shading.}
\label{fig:ElemEvents}
\end{figure}
CGF in Eq.~\eqref{eq:S1} accounts for
the Andreev reflection of the excess electrons due to dc voltage
applied, cf. Fig.~\ref{fig:ElemEvents}(a). The charges are transferred
in pairs and the statistics is binomial in each transport channel
with the transmission probabilities given by the Andreev reflection
coefficients $R^A_n$. The rate $2e\Vdc/h$ with which the excess electrons
impinge on the contact is the same as in the normal case.
CGF in Eq.~\eqref{eq:Seh} accounts for the charge transfer statistics
due to electron-hole pairs created by the ac drive.
In general, $\cS_{eh}$ consists of two types of electron-hole
processes with different probabilities $p_k$ of the electron-hole
pair creations and different numbers of attempts for particles to
traverse the junction,
$M=(\hbar\omega t_0/h)(1-v)$ and $M=(\hbar\omega t_0/h)v$.
Here, $v=2e\Vdc/\hbar\omega - \lfloor 2e\Vdc/\hbar\omega \rfloor$
is a fractional part of $2e\Vdc/\hbar\omega$
($\lfloor x \rfloor$ denotes the integer part of $x$).
Charge transfers described by Eq.~\eqref{eq:Seh} are bidirectional
processes in which an electron from the pair is Andreev reflected
at the contact while the hole exhibits a normal reflection,
or vice versa. This is schematically depicted in
Fig.~\ref{fig:ElemEvents}(b). As a result, pairs of charge quanta are
transferred in either direction with the probabilities
$R^A_n(1-R^A_n)p_k$ in each transport channel.

\section{Andreev levitons in an NS junction}
\label{sec:ex}

\begin{figure}[t]
\begin{center}
\includegraphics[scale=1.45]{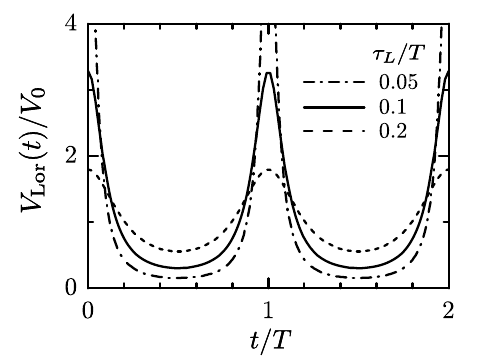}
\end{center}
\caption{Lorentzian voltage pulses for different widths:
$\tau_L/T=0.05$ (dash-dotted), $0.1$ (solid), and $0.2$
(dashed line).}
\label{fig:Voltage}
\end{figure}

As a first example we compute the photon-assisted noise when the junction is
driven by periodic Lorentzian voltage pulses of the form:
\begin{align}
V_{\rm Lor}(t)
& =
\frac{V_0}{\pi} \sum_{k=-\infty}^\infty
\frac{T\tau_L}{(t-kT)^2 + \tau_L^2} \notag
\\
&=
\frac{V_0 \sinh(2\pi\tau_L/T)}
{\cosh(2\pi\tau_L/T) - \cos(2\pi t/T)}.
\end{align}
Here, $T=2\pi/\omega$ is the period of the drive, and the
amplitude $V_0$ is chosen in such a way as to represent the
average voltage per period,
$(1/T) \int_0^T V_{\rm Lor}(t)dt = V_0$.
Lorentzian voltage pulses are shown in Fig.~\ref{fig:Voltage}
for different widths $\tau_L/T = 0.05$, $0.1$, $0.2$.

The excess photon-assisted noise (the total noise minus dc noise
level) is given by
\begin{equation}
\frac{S_{\rm ac}}{S_0} = \sum_{n=-\infty}^\infty
\left\vert \frac{2eV_0}{\hbar\omega} + n \right\vert\, |a_n|^2
- \left\vert \frac{2eV_0}{\hbar\omega} \right\vert,
\end{equation}
where
\begin{equation}\label{eq:S0}
S_0 = \frac{4e^2\hbar\omega}{h} \sum_n R^A_n(1-R^A_n).
\end{equation}
Coefficients $a_n$ are related to the doubled ac part of the
drive voltage, $2V_{\rm ac}(t) = 2(V_{\rm Lor}(t) - V_0)$:
\begin{equation}
a_n = \frac{1}{T} \int_0^T dt\,
e^{-i\phi(t)}\, e^{in\omega t},
\end{equation}
where $\phi(t)=(e/\hbar)\int_0^t 2V_{\rm ac}(t')dt'$.
These coefficients read \cite{dubois_integer_2013}:
\begin{equation}
a_n = q \gamma^n \sum_{k=0}^\infty
\frac{(-1)^k\, \gamma^{2k}\, \Gamma(q+n+k)}
{\Gamma(k+1)\Gamma(q-k+1)\Gamma(n+k+1)}
\end{equation}
for $n \ge 0$, and
\begin{equation}
a_n = q \gamma^{|n|} \sum_{k=0}^\infty
\frac{(-1)^{k+n}\, \gamma^{2k}\, \Gamma(q+k)}
{\Gamma(k+1) \Gamma(q-|n|-k+1) \Gamma(|n|+k+1)}
\end{equation}
for $n<0$. Here, $q=2eV_0/\hbar\omega$ and $\gamma = e^{-2\pi\tau_L/T}$.

\begin{figure}[t]
  \centering
  \includegraphics[width=0.9\columnwidth]{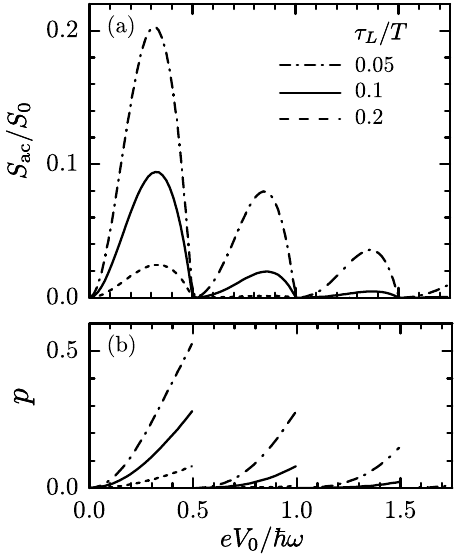}
  \caption{(a) Excess photon-assisted noise due to a Lorentzian
  voltage drive shown in Fig.~\ref{fig:Voltage} and (b) the
  corresponding probabilities of the electron-hole pair creations as
  a function of the drive amplitude.} 
  \label{fig:NoisePk3}
\end{figure}

The photon-assisted noise is shown in Fig.~\ref{fig:NoisePk3}(a).
As the width $\tau_L$ is increased, the pulses overlap more strongly
and $V_{\rm Lor}(t)$ approaches the constant voltage $V_0$. This
results in the overall suppression of the excess photon-assisted
noise. In addition, the excess noise is fully suppressed at
half-integer values of $eV_0/\hbar\omega$. This corresponds to
the half-integer Lorentzian pulses $e\int V_{\rm Lor}(t)dt = nh/2$,
in contrast to the normal junctions in which the noise suppression
occurs for integer pulses.
%
%
%
The photon-assisted noise can also be expressed in terms of
elementary events of the electron-hole pair creations
\cite{VanevicNazarovBelzigPRL99-07,VanevicNazarovBelzigPRB78-08}:
\begin{equation}
S_{\rm ac}/S_0=2(1-v)\sum_k p_k.
\end{equation}
Here, $v=2eV_0/\hbar\omega - \lfloor 2eV_0/\hbar\omega \rfloor$
is the fractional part of $2eV_0/\hbar\omega$
and $p_k$ are the probabilities of the pair creations given by
$p_k = \sin^2(\alpha_k/2)$ where $e^{\pm i \alpha_k}$ are the
eigenvalues of the operator $h\tilde h$.
The electron-hole creation probabilities as a function
of the amplitude of the Lorentzian pulses are shown
in Fig.~\ref{fig:NoisePk3}(b). We find that in the problem at hand,
there is only one electron-hole pair created per period with
probability $p_1\equiv p$. The pair creation probability
increases as $eV_0/\hbar\omega$ approaches the half-integer values.
However, the photon-assisted noise is nevertheless
zero at these points because the effective rate of
attempts $\hbar\omega(1-v)/h$ vanishes.

\section{Large excitation noise}
\label{sec:largevoltage}

In what follows we allow for the dc bias and the ac drive amplitudes
to be comparable to $\Delta$. For simplicity, we still
assume low-temperature limit, $T_e=0$.

When the drive amplitudes are comparable to $\Delta$, the
NS junction can no longer be mapped to the normal one. However, we
can proceed with the numerical calculation of the
cumulant generating function in Eq. \eqref{eq:S-a}.
The Green's functions are given by
\begin{align}
\vG_N(0) &=
\begin{pmatrix}
    \bar\tau_3 & 2\bar h \\
    0 & -\bar\tau_3
\end{pmatrix},
\\
\vG_S(0) &=
\begin{pmatrix}
    \bar G_R & (\bar G_R - \bar G_A)h(\cE) \\
     0       & \bar G_A
\end{pmatrix}.
\end{align}
Here,
\begin{equation}
\bar h =
\begin{pmatrix}
    h_1 &  0 \\
     0  & -h_2
\end{pmatrix},
\end{equation}
\begin{equation}
\bar G_{R,A} =
\frac{\pm 1}{\sqrt{(\cE\pm i0)^2 - \Delta^2}}
\begin{pmatrix}
    \cE\pm i0 & \Delta \\
    -\Delta & -(\cE\pm i0)
\end{pmatrix},
\end{equation}
and $h_1=U h U^\dagger$, $h_2 = U^\dagger h U$.

Next, we note that for the periodic time-dependent drive
with the period $T=2\pi/\omega$, the operators $h_{1,2}$
couple only the energies that differ by an integer multiple of
$\hbar\omega$. Therefore we can use a matrix representation in
energy indices,
$(h_i)_{nm}(\epsilon) = h_i(\epsilon+n\hbar\omega,\epsilon + m\hbar\omega)$
($-\hbar\omega/2<\epsilon<\hbar\omega/2$).
This provides a matrix structure in energy in $\vG_N$ and $\vG_S$.
The trace operation in Eq.~\eqref{eq:S-a} now amounts to a matrix
diagonalization in Keldysh, Nambu, and energy indices and integration
over $\epsilon$.

We calculate the cumulant generating function for a diffusive
NS junction that has a distribution of transmission eigenvalues
given by
\begin{equation}
\rho(T) = \frac{G_N}{G_Q} \frac{1}{2T\sqrt{1-T}},
\end{equation}
where $G_N$ is the normal-state conductance and $G_Q=2e^2/h$ and it is
assumed that the Thouless energy is much larger than all other
relevant energy scales.
The cumulant generating function in Eq. \eqref{eq:S-a} reduces to
\begin{equation}
\cS(\chi) = -\frac{G_N}{2G_Q}\Tr
\arcsin^2 \left(\frac{1}{2} \sqrt{2-\acom{\vG_N(\chi)}{\vG_S}} \right).
\end{equation}
For the average current and the current noise power we obtain
\begin{equation}
I = G_N \int_{-\hbar\omega/2}^{\hbar\omega/2} \frac{d\epsilon}{4e}
\sum_\lambda
\frac{\partial\arcsin^2(\sqrt{\lambda(\epsilon,\chi)}/2)}
{\partial (i\chi)}\Big|_0,
\end{equation}
\begin{equation}
S = - G_N \int_{-\hbar\omega/2}^{\hbar\omega/2} \frac{d\epsilon}{4}
\sum_\lambda
\frac{\partial^2\arcsin^2(\sqrt{\lambda(\epsilon,\chi)}/2)}
{\partial (i\chi)^2}\Big|_0,
\end{equation}
where $\lambda$ are the eigenvalues of $2-\acom{\vG_N(\chi)}{\vG_S}$.

\begin{figure}[t]
\centering
\includegraphics[width=0.9\columnwidth]{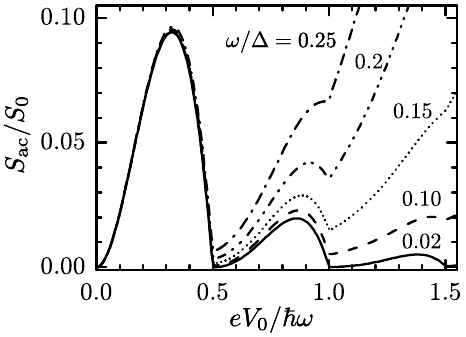}
\caption{Excess photon-assisted noise for the Lorentzian pulses
with $\tau_L/T=0.1$. The noise is shown for different drive
frequencies: $\hbar\omega/\Delta=0.25$, $0.20$, $0.15$, $0.10$, $0.02$
(top to bottom).}
\label{fig:acNoiseLorW1}
\end{figure}

Photon-assisted noise $S_{\rm ac}=S - S_{\rm dc}$ for an NS
junction driven by periodic Lorentzian pulses is shown in
Fig. \ref{fig:acNoiseLorW1} for different drive frequencies
$\hbar\omega/\Delta=0.25$, $0.2$, $0.15$, $0.1$, $0.02$ (top to bottom).
The noise is normalized to $S_0$ in Eq. \eqref{eq:S0}
which in the case of a diffusive junction reads
$S_0 = G_N\hbar\omega/3$.

The half-integer Lorentzian pulses with $eV_0/\hbar\omega=n/2$
at energies much smaller than the superconducting gap
create Andreev levitons which are the minimal excitation states
of the NS system. As the energy becomes comparable to the gap,
the system can no longer be mapped to an effective normal junction.
Therefore, the half-integer Lorentzian pulses are no longer
optimal due to contributions of normal and Andreev electron transport
above the gap. As a result, the excess photon assisted noise starts
to increase, see Fig. \ref{fig:acNoiseLorW1}. However,
at energies much larger than the gap, the junction is in the normal
state and the {\em integer} Lorentizan pulses create minimal excitation
states in the normal junction.

\begin{figure}[t]
\centering
\includegraphics[width=0.9\columnwidth]{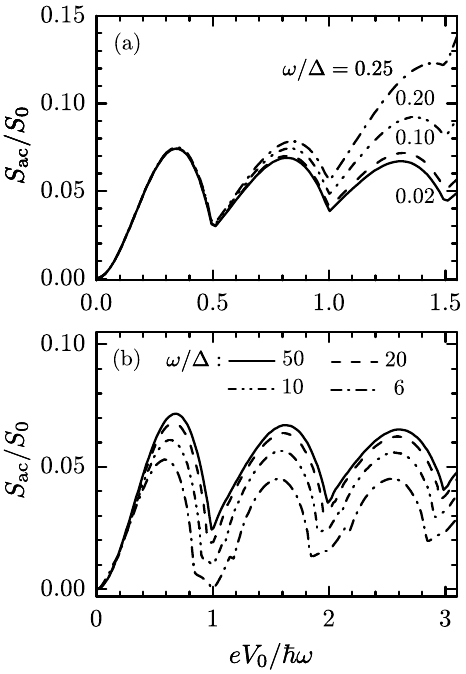}
\caption{Excess photon-assisted noise for the harmonic time-dependent
voltage $V(t)=V_0[1+\cos(\omega t)]$ and amplitudes of the drive
that are (a) smaller and (b) larger than the gap.
At subgap energies the excess noise is minimal at half-integer
drive amplitudes $eV_0/\hbar\omega = n/2$, while above the gap it
is minimal at integer drive amplitudes $eV_0/\hbar\omega = n$.}
\label{fig:acNoiseCos}
\end{figure}

The quantum oscillations of the photon assisted noise
as a function of the dc voltage have been observed recently
in the normal-state tunnel junction driven by harmonic time-dependent
voltage $V(t) = V_0[1+\cos(\omega t)]$ \cite{ReuletPRB88-13}.
Harmonic drive in general creates additional electron-hole pairs
which results in the non-zero excess noise with minima
at integer values $eV_0/\hbar\omega=n$.
The same is true in the NS junction at energies much lower
than the gap, except that the excess noise minima appear
at half-integer values $eV_0/\hbar\omega=n/2$, see
Fig. \ref{fig:acNoiseCos}. At intermediate drive frequencies
that are comparable to the gap, the noise in the system is
determined by a density of states which is affected both
by the drive and by the superconducting proximity effect.
In addition, both normal and Andreev processes contribute
to the transport. As a result, the total noise can even
be suppressed below the effective dc noise level
(the excess noise can be negative), the situation which
otherwise cannot occur in the normal junction.


\section{Conclusion}
\label{sec:conclusion}

We have analyzed the transport properties of a driven quantum point
contact between a normal metal and a superconductor. Using an extended
Keldysh Greens function method we could determine the full counting
statistics and identify the elementary transport processes for an
arbitrary voltage drive in the subgap regime. At voltage amplitudes
and frequencies well below the superconducting gap $\Delta$ the
NS-contact can be mapped onto a contact with two normal leads. The
transmission is determined by the Andreev reflection probability
and the probabilities of elementary Andreev processes are determined by
the ones known for normal transport with an effective charge $2e$. In
that spirit we have discussed Lorentzian voltage pulse which lead to
Andreev-Levitons, which are now pure two-charge excitations, for
half-integer quantized amplitudes $e\int dt V=nh/2$. Finally we have
discussed the transition from half-integer steps
$eV_0/\hbar\omega=n/2$ in the ac-noise to simply quantized steps for 
voltages and frequencies much larger than $\Delta$.

In future it will be interesting to investigate open questions, e.g.,
for which parameters the noise is minimized and the nature of the
elementary events at intermediate frequencies where both Andreev 
and normal reflection processes coexist, as well as the effects of
dephasing related to the finite Thouless energy.   

We acknowledge financial support by DFG through SFB 767 and BE3803/5.
MV acknowledges the Serbian Ministry of Science Project No. 171027.








\end{document}